\documentclass[twocolumn,showpacs,prb]{revtex4}
\usepackage{graphicx}
\usepackage{bbold}
\usepackage{color}
\usepackage[normalem]{ulem}

\begin{document}

\title{
Dynamical Magnetic and Nuclear Polarization in Complex Spin Systems: Semi-magnetic II-VI Quantum Dots}

\author{Ramin M. Abolfath$^{1,2,3,\dagger}$, Anna Trojnar$^{2,3}$, Bahman Roostaei$^4$, Thomas Brabec$^{2}$, Pawel Hawrylak$^{3}$}

\affiliation{
$^{1}$School of Natural Sciences and Mathematics, University of Texas at Dallas, Richardson, TX 75080 \\
$^{2}$University of Ottawa, Physics Department 150 Louis Pasteur, Ottawa, ON, K1N 6N5, Canada \\
$^{3}$Institute for Microstructural Sciences, National Research Council of Canada, Ottawa, Ontario, Canada K1A~0R6 \\
$^{4}$Institut f\"ur Theoretische Physik, Universit\"at zu K\"oln, K\"oln, Germany
}
\date{\today}

\begin{abstract}
Dynamical magnetic and nuclear polarization in complex spin systems is discussed on the example of transfer of spin from exciton to the central spin of magnetic impurity in a quantum dot in the presence of a finite number of nuclear spins. The exciton is described in terms of the electron and heavy hole spins  interacting via exchange interaction with  magnetic impurity, via hyperfine interaction with a finite number of nuclear spins and via dipole interaction with photons.  The  time-evolution of the exciton, magnetic impurity and nuclear spins is calculated exactly between quantum jumps corresponding to exciton radiative recombination.
The collapse of the wavefunction and the refilling of the quantum dot with new spin polarized exciton is shown to lead to build up of magnetization of the magnetic impurity as well as nuclear spin polarization. The competition between electron spin transfer to magnetic impurity and to nuclear spins simultaneous with the creation of dark excitons is elucidated. The technique presented here opens up the possibility of studying optically induced Dynamical Magnetic and Nuclear Polarization in Complex Spin Systems.
\end{abstract}
\pacs{75.50.-y,75.50.Pp,85.75.-d}

\maketitle

\section{Introduction}
\label{sec0}
There is currently interest in developing means of localizing and controlling complex spin systems in solid state devices~\cite{Hanson2007:RMP}. This includes electron and/or hole spins in gated~\cite{Pioro-Ladriere2003:PRL,Petta2005:S}, self-assembled~\cite{Bayer2001:S}, nanocrystal~\cite{Ochsenbein2009:NN} and carbon nanotube quantum dots~\cite{Churchill2009:PRL}, nitrogen vacancies in diamond~\cite{Bassett2011:PRL} and magnetic impurities in II-VI~\cite{Goryca2009:PRL,LeGall2009:PRL,Besomebes2009:SSC,Mackowski2004:APL,Gould2006:PRL,Klopotowski2011:PRB} and III-V~\cite{Viswanatha2011:PRL,Baudin2011:PRL} semiconductors. The complex spin systems involved include heavy valence holes with spin $J=3/2$, nitrogen vacancies with spin $M=1$, half-filled shell electrons of mangan $Mn^{2+}$ impurity atom with spin $M=5/2$ or $Mn^{3+}$ atom with $M=3/2$ in II-VI semiconductor quantum dots, or strongly coupled valence hole-Mn atom in InAs/GaAs quantum dots.
Extensive theoretical studies have been carried out, predicting rich phase diagram for these systems~\cite{Fernandez2004:PRL,Govorov2004:PRB,Qu2006:PRL,Abolfath2008:PRL,Cheng2008:EPL}.
For  NV centers in diamond, carbon nanotube based quantum dots and magnetic impurities in II-VI semiconductor quantum dots the complex spin system  interacts with only a finite  number of nuclear spins.
The controlling of magnetization of complex spin systems is often carried out optically and involves transfer of photon angular momentum into exciton spin, and exciton spin into the spin of the complex spin system~\cite{Kirilyuk2010:RMP,Oiwa2002:PRL}.
This dynamical magnetic polarization (DMP) process is decohered by photon and nuclear spin baths.
Recently, first optical experiments on single magnetic impurities in II-VI quantum dots measured the dynamic evolution of the magnetization process~\cite{Goryca2009:PRL,LeGall2009:PRL,Besomebes2009:SSC} with theoretical models of DMP based on rate equations~\cite{Reiter2009:PRL,Cywinski2010:PRB}.

In this work, we develop a  microscopic theory of optically driven dynamical magnetic polarization of complex spin systems.
The theory describes the transfer of spin from exciton to the central spin of magnetic impurity in a quantum dot in the presence of a finite number of nuclear spins using quantum jump approach~\cite{Plenio1999:RMP,Hohenester2001:SSC,ScullyBook}. The exciton is described in terms of
the electron and heavy hole spins  interacting via exchange interaction with  magnetic impurity, via hyperfine interaction with a finite number of nuclear spins and via dipole interaction with photons.  The  time-evolution of the exciton, magnetic impurity and nuclear spins is calculated exactly between quantum jumps corresponding to exciton radiative recombination.
The collapse of the wavefunction and the refilling of the quantum dot with new spin polarized exciton as in recent experiment by Goryca et al.~\cite{Goryca2009:PRL} is shown to lead to build up of magnetization of the magnetic impurity as well as nuclear spin polarization. The competition between electron spin transfer to magnetic impurity and to nuclear spins simultaneous with the creation of dark excitons is elucidated.

The paper is organized as follows.
In Section~\ref{sec1} we describe our model.
Section~\ref{sec2} describes quantum jump approach to time evolution of a single MI and a single exciton 
in the absence of nuclear spins.
Section~\ref{sec4} contains quantum jump approach and the dynamical evolution of MI interacting with a train of excitons in the presence of nuclear spins.
In sections~\ref{sec5} and~\ref{sec6} we present numerical results, discussions, conclusion and the
summary.

\section{The model}
\label{sec1}
We consider a semiconductor QD containing  a  complex spin system $M$, e.g., magnetic ion (MI), coupled with  few nuclear spins of the host material, as in, e.g., CdTe quantum dots. The quantum dot with MI is attached to a smaller quantum dot with no MI where the electrons and valence holes with definite spin polarization are generated optically by circularly polarized light. This is illustrated in Fig.1a where circles describe quantum dots, blue arrow corresponds to electron spin $S_{z}=+1/2$ and  white arrow corresponds to heavy hole spin  $J_{z}=- 3/2$ in the smaller dot.
The larger dot contains a randomly oriented complex spin $M$, represented by a magenta arrow and a number of randomly oriented nuclear spins represented by small arrows. The DMP process starts with transfer of spin polarized exciton from the smaller QD to the larger QD, as illustrated in Fig.1b. As a result of interactions, the spin of electron, MI and nuclear spins undergo a flip-flop process as the wavefunction of the larger dot evolves and forms an entangled state,  a linear combination of bright and dark excitons, as shown in Fig.1c. During this process the smaller dot is refilled with spin polarized exciton. 
Simultaneously, the bright exciton decays due to interaction with the photon field with a random recombination time, resulting in a photon emission and a quantum jump takes place.
As a result of this process, the states of the magnetic ion and nuclear spins are modified, the polarization is increased and the larger dot is refilled with spin polarized exciton and the DMP  process continues.

\begin{figure}
\begin{center}
\vspace{1cm}
\includegraphics[width=0.98\linewidth]{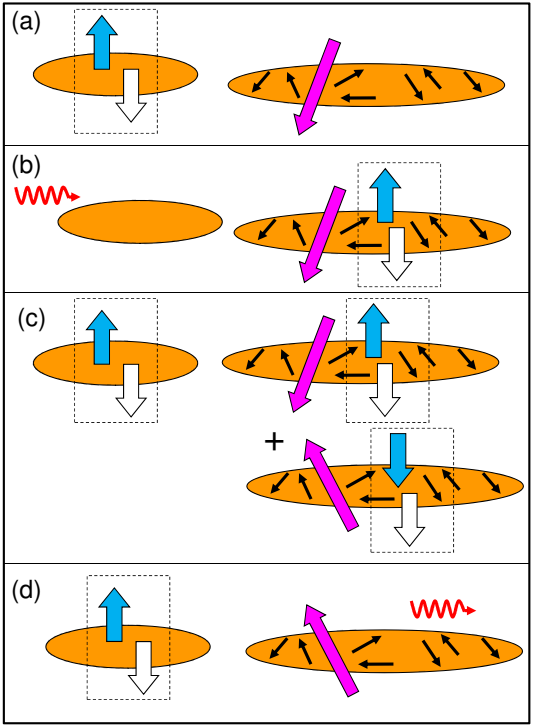}
\caption{Schematic representation of the DMP process.}
\label{fig1}
\end{center}
\end{figure}

We now quantitatively describe the DMP process. We start with the Hamiltonian  describing the quantum dot coupled with the photon bath  $H =   H_{\rm QD}+ H_{\rm ph QD}+ H_{\rm ph }$ . Here $H_{\rm ph }$ is the photon Hamiltonian, $H_{\rm ph QD }$ is the Hamiltonian describing coupling of photons with the exciton in a QD and $H_{QD}$ is the QD hamiltonian.

The QD Hamiltonian describes the coupling between exciton $X$ and the magnetic moment of the complex spin system, consisting of MI and nuclear spins $I$. It is given by: $H_{QD} =  H_{\rm m} + H_{\rm x} + H_{\rm xm} +  H_{\rm n} + H_{\rm xn} + H_{\rm mn}$.
Here $H_{\rm x}$ describes the exciton internal energy, $H_{\rm m}$ describes the MI internal energy and the remaining terms in $H_{QD}$ represent X-MI, X-I and MI-I exchange couplings.
The exciton Hamiltonian describes the
low energy quadruplet $|S,J\rangle$ characterized by quantum spin numbers of an electron, $S=\pm 1/2$, and a heavy hole, $J= \pm 3/2$ in the QD. The complex spin system is described by a total spin $\vec{M}=\sum_{i=1}^{N} \vec{u}_{i}$, where $N$ is the number of spins $u=1/2$ building up the MI system, and
$H_m = \sum_{i<j} J_{ij} \vec{u}_{i} \cdot \vec{u}_{j} + D M_z^2$ in which $J_{ij}$ are exchange matrix elements building the total spin $M$. In quantum dots one often includes strain field
$D$ leading to splitting of the different $M_z$ levels~\cite{Furdyna1988:JAP}.
Similarly $\vec{I} = \sum_{i=1}^{N_b} \vec{I}_i$ where $N_b$ is total number of nuclear spins.

We assume that exchange coupling constants of MI spins with the environment  are identical and
 the full QD Hamiltonian can be written as~\cite{Cheng2008:EPL}:
\begin{eqnarray}
&&H_{QD} = H_{\rm m} + H_{\rm x} + J_{\rm hm} J_{z} M_z - J_{\rm em} \vec{S} \cdot \vec{M} \nonumber \\ &&
+ \sum_{\rm n}^{\rm N_b} [J_{\rm ne}\vec{I}_{\rm n}\cdot\vec{S} + J_{\rm nh} I_{\rm z,n} J_z]
+ \sum_{\rm n}^{\rm N_b} \sum_{\rm n'\neq n}^{\rm N_b} J_{\rm nn'}\vec{I}_{\rm n}\cdot\vec{I}_{\rm n'} \nonumber \\ &&
+ \sum_{\rm n}^{\rm N_b} A_n \vec{I}_{\rm n}\cdot\vec{M}.
\end{eqnarray}
The exciton Hamiltonian  $H_{\rm x}=\Delta_0 S_z J_z + \Delta_1 ( S^+ J^- + S^- J^+)$ describes splitting $\Delta_0$
between the low energy dark exciton doublet $|\uparrow,\Uparrow\rangle=|+1/2,+3/2\rangle,~|\downarrow,\Downarrow\rangle=|-1/2,-3/2\rangle$ with total angular momentum $j_z=\pm 2$ along quantization-axis, $\hat{z}$, and higher energy bright exciton doublet $|\downarrow,\Uparrow\rangle=|-1/2,+3/2\rangle,~|\uparrow,\Downarrow\rangle=|-3/2,+1/2\rangle$ with $j_z=\pm 1$. Here $\uparrow/\downarrow$ and $\Uparrow/\Downarrow$ represent spin of electron and hole~\cite{Bayer2002:PRB}. The bright exciton doublet is split by
the anisotropic electron-hole exchange interaction characterized by parameter $\Delta_1$ which measures the splitting of the two bright exciton states $|+1/2,-3/2\rangle,~|-1/2,+3/2\rangle$. $\Delta_1$ is zero for cylindrical quantum dots and the two bright exciton states correspond to circular photon polarization.
The exciton-MI coupling in Eq.1 is given as a sum of the ferromagnetic Heisenberg electron-MI exchange $H_{\rm em} = - J_{\rm em} \vec{S}\cdot\vec{M}$ and anti-ferromagnetic Ising exchange interaction $H_{\rm hM} = + J_{\rm hM} J_z M_z$.~\cite{Cheng2008:EPL}
Only electron-MI interaction is responsible for the  e-MI spin flip-flop process.
The interaction of complex spin MI with nuclear spin associated with the spin complex is denoted here by $H_{MI}=A \vec{I}_M \cdot \vec{M}$. This interaction might, for example, describe coupling of manganese d-shell electron spins with manganese ion nuclear spin~\cite{Furdyna1988:JAP}. With hole spin strongly aligned along the growth $z$ direction the coupling  of electron and hole spins to surrounding nuclear spins of isotopes of the QD and barrier material with finite nuclear spin reads $\sum_{n}^{N_b} [J_{\rm ne}\vec{I}_{\rm n}\cdot\vec{S} + J_{\rm nh} I_{\rm z,n} J_z]
+ \sum_{\rm n,n'} J_{\rm nn'}\vec{I}_{\rm n}\cdot\vec{I}_{\rm n'}$ where $N_b$ is the number of nuclear spins in the QD and the last term describes nuclear spin interaction.
We note that for isotropic QD the long range e-h exchange $\Delta_1$ is zero and the heavy hole spin $J_{z}=\pm 3/2$ is preserved.

\section{Single MI, single exciton and no nuclear spin}
\label{sec2}
We start our discussion of DMP by discussing time evolution of magnetization of  X-MI complex interacting with harmonic fields of photons in the absence of nuclear spins.
To focus on quantum dynamics in the simplest spin system, we consider  MI with $M=1/2$ and just two states, $|\uparrow\rangle=|M_z=1/2\rangle$ and $|\downarrow\rangle=|M_z=-1/2\rangle$, and Hamiltonian $H_{QD} = H_{\rm x} + H_{\rm xm}$ where $H_{\rm xm} = - J_{\rm em} \vec{S}_e\cdot \vec{M} + J_{\rm hm} S_{z,h} M_z$.
We also consider a CW laser field with one type of circular polarization, e.g., $\sigma=+1$, that generates excitons with one type of polarization, $j_z=+1$, corresponding to $|X_b\rangle = |\downarrow,\Uparrow\rangle$.
Because we neglect the hole spin-flip in the spin flip-flop process of X-MI complex, as discussed in Sec.~\ref{sec1}, the dark exciton $|X_d\rangle = |\uparrow,\Uparrow\rangle$ with $j_z=+2$ is the only state generated throughout the electron-MI spin-flip.
Hence the space of a single X-MI complex can be spanned by $|1\rangle = |X_b,\downarrow\rangle$,  $|2\rangle = |X_b,\uparrow\rangle$, $|3\rangle = |X_d,\downarrow\rangle$,  $|4\rangle = |X_d,\uparrow\rangle$, $|5\rangle = |0,\downarrow\rangle$,  $|6\rangle = |0,\uparrow\rangle$.
In this basis the exciton Hamiltonian is diagonal $H_{\rm x}={\rm diag}(E_{b}, E_{b}, E_{d}, E_{d}, 0, 0)$.
Here $E_b$, and $E_d$ are the energy of bright and dark excitons measured relative to the vacuum.
In the basis of $\{|1\rangle, |4\rangle\}$, $\{|2\rangle, |3\rangle\}$, and $\{|5\rangle, |6\rangle\}$, the X-MI Hamiltonian is block-diagonal
$H_{\rm xm}=H_{\rm xm,1}\oplus H_{\rm xm,2}\oplus H_{\rm xm,3}$ where $H_{\rm xm,1}=(-J_{\rm em}+J_{\rm hm})/4 ~\mathbb{1}$,
$H_{\rm xm,2} =
\left(\begin{array}{cc}
(J_{\rm em}-J_{\rm hm})/4 & -J_{\rm em}/2 \\
-J_{\rm em}/2 & (J_{\rm em}-J_{\rm hm})/4
\end{array}\right)$,
and $H_{\rm xm,3}=0$ respectively.
Here $\mathbb{1}$ is a $2\times 2$ unit matrix.

\begin{figure}
\begin{center}\vspace{1cm}
\includegraphics[width=1.0\linewidth]{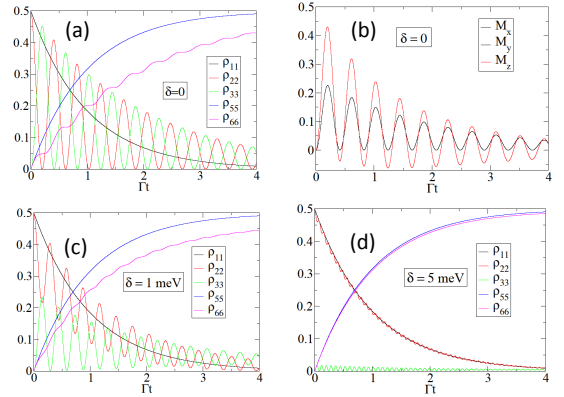}\\ \vspace{-.5cm}
\caption{
Time evolution of density matrix with the initial condition $\rho_{11}(t=0)=\rho_{22}(t=0)=0.5$ and $\rho_{ij}(t=0)=0$ for other $i$ and $j$'s, corresponding to initial random state of MI, are shown in (a), (c), and (d) for $\delta=0,1,5$ meV. In (b) the expectation value of MI spin for $\delta=0$ is shown. Here $|1\rangle = |X_b,\downarrow\rangle$,  $|2\rangle = |X_b,\uparrow\rangle$, $|3\rangle = |X_d,\downarrow\rangle$,  $|4\rangle = |X_d,\uparrow\rangle$, $|5\rangle = |0,\downarrow\rangle$,  $|6\rangle = |0,\uparrow\rangle$. The population of vacuum can be calculated by $\rho_{\rm vacuum} = \rho_{55} + \rho_{66}$. The elements of density matrix, not plotted in this figure, are all identical to zero.}
\label{fig0}
\end{center}
\end{figure}
The off-diagonal elements of $H_{\rm xm,2}$ describe mixing of $X_b$ and $X_d$ via spin-1/2 MI.
Hence
$|\psi(t)\rangle = C_{b,\uparrow}(t)\exp(-i E_b t/\hbar) |X_b, \uparrow\rangle + C_{d,\downarrow}(t) \exp(-i E_d t/\hbar) |X_d,\downarrow\rangle$
with initial condition $|\psi(t=0)\rangle = |X_b, \uparrow\rangle$ is one of the solutions of the time-dependent Schr\"odinger equation $-i\hbar\frac{\partial}{\partial t} |\psi(t)\rangle = H_{QD} |\psi(t)\rangle$.
This state describes a coherent Rabi-oscillations between bright and dark excitons due to MI spin flip-flop.

Note that in $|\psi(t)\rangle$ there is no mixing with the vacuum, $|0, M_z=\pm 1/2\rangle$, unless we take into account the coupling of bright-exciton with radiation field.
In the interaction and rotating wave approximation the electron-photon coupling is described by the Hamiltonian
that does not directly change the state of MI
\begin{eqnarray}
H_{\rm ph QD}(t)&=&\hbar\sum_{\vec k,M_z} g_{\vec k} [b^\dagger_{\vec k} |0,M_z\rangle \langle X_b,M_z| e^{-i(\omega_b - \omega_{\vec k})t} \nonumber \\ &&
+ b_{\vec k} |X_b,M_z\rangle \langle 0,M_z| e^{+i(\omega_b - \omega_{\vec k})t}],
\label{eq0001}
\end{eqnarray}
where $b^\dagger_{\vec k}$ and $b_{\vec k}$ are creation and annihilation operators of photon with specific circular polarization $\sigma=+1$. $g_{\vec k}$, and $\omega_{\vec k}$ are the photon-X coupling constant and photon frequency, respectively, and $\omega_b=E_b/\hbar$.
The equation of motion of the QD density matrix, $\rho$, coupled with thermal bath of photons can be calculated after tracing over photon degrees of freedom.
Here $\rho$ represents the density matrix of a single exciton interacting with a single MI.
Assuming that photons are in thermal equilibrium and are weakly coupled with excitons in QDs, the equation of motion for exciton density matrix, $\rho$, can be calculated perturbatively. Up to the second order of perturbation, it is straightforward to show that~\cite{ScullyBook}
\begin{eqnarray}
\frac{d\rho}{dt} &=& -\frac{i}{\hbar}[H_{QD},\rho] -\frac{\Gamma}{2} n_B \sum_{M_z}
(|0,M_z\rangle \langle 0,M_z| \rho \nonumber \\ &&
- 2 |X_b,M_z\rangle \langle 0,M_z| \rho |0,M_z\rangle \langle X_b,M_z| \nonumber \\ &&
+ \rho |0,M_z\rangle \langle 0,M_z|) \nonumber \\ &&
-\frac{\Gamma}{2} (n_B + 1) \sum_{M_z}
(|X_b,M_z\rangle \langle X_b,M_z| \rho \nonumber \\ &&
- 2 |0,M_z\rangle \langle X_b,M_z| \rho  |X_b,M_z\rangle \langle 0,M_z| \nonumber \\ &&
+ \rho |X_b,M_z\rangle \langle X_b,M_z|),
\label{eq02}
\end{eqnarray}
where $n_B = 1/(e^{\hbar\omega/k_BT}-1)$ is Bose-Einstein distribution function and
$\Gamma=\frac{4\omega_b^3 \xi^2}{3\hbar c^3}$ is the transition rate for the spontaneous emission of photons. $\xi$ is the dipole moment matrix element.
Note that in Eq.(\ref{eq02}) vacuum can be considered as a shelving-state.

\begin{figure}
\begin{center}
\vspace{1cm}
\includegraphics[width=0.98\linewidth]{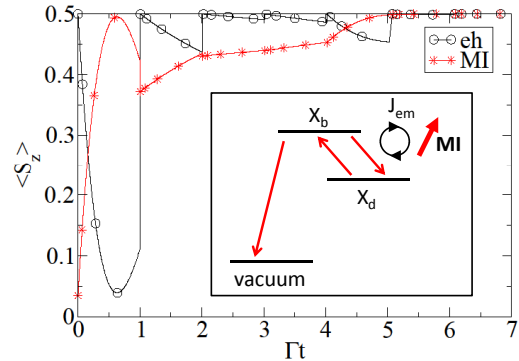}
\caption{
Time evolution of $S_z$ of a train of injected photo-electrons inside QD (circles) and a single MI (stars) with spin $S=1/2$ and no nuclear spin.
The inset shows the $\Lambda$-shape three level optical resonance of bright- and dark-exciton ($X_b$ and $X_d$). The Rabbi oscillation between $X_b$ and $X_d$ occurs because of the exchange interaction between the exciton and the system of MI and nuclear spins.
The optical selection rule allows decay of $X_b$ to vacuum. However, the population of $X_d$ decreases indirectly through the conversion of $X_d$ to $X_b$.
}
\label{fig3}
\end{center}
\end{figure}

The numerical solutions of Eq.(\ref{eq02}) at zero-temperature ($n_B=0$) are shown in Fig.~\ref{fig0} for a QD with $E_d=2$ eV and $\delta = E_b - E_d = 0, 1, 5$ meV. Here we used $J_{\rm em}=1$ meV and $J_{\rm hm}=4$ meV.
The initial state of MI is completely uncorrelated with half of the spins populated in up-direction.
As it is shown, because of the coupling with the bath of photons, bright-exciton decays to vacuum without flipping the MI spin and mixing with $X_d$, e.g., $|X_b, M_z\rangle \rightarrow |0, M_z\rangle$.
However, a coherent Rabi oscillation between $X_b$ and $X_d$ via exchange with MI is responsible for spin-transfer to MI.
In Fig.~\ref{fig0}(b) the time evolution of the components of the ensemble-averaged magnetization of MI, $\langle M_\alpha\rangle = {\rm Tr} (\rho M_\alpha)$ with $\alpha=x,y,z$ are depicted for $\delta=0$.
As it is shown, $\langle \vec{M}(t)\rangle$ exhibits under-damped oscillations around a positive field that decays to zero as a function of time.
In Fig.~\ref{fig0}, we find that $\rho_{11}=e^{-\Gamma t}/2$ and $\rho_{55}=(1-e^{-\Gamma t})/2$ fit perfectly the numerical solution of $\rho_{11}$ and $\rho_{55}$ for all $\delta$s.
The decay channel of dark-exciton is through a transition to bright-exciton and spin-flip of MI.
This process is schematically depicted in the inset of Fig.~\ref{fig3}.
A strong dependence of dark-exciton population on $\delta$ is seen in Fig.~\ref{fig0}.

Consistent with the time-evolution of the density matrix, we propose an exciton wave-function that fits the density matrix via $\rho(t)=|\psi(t)\rangle \langle \psi(t)|$:
\begin{eqnarray}
&&|\psi(t)\rangle = C_{b\downarrow} e^{-\Gamma t/2} |X_b, \downarrow\rangle
+ C_{b\downarrow} \sqrt{1 - e^{-\Gamma t}} |0, \downarrow\rangle \nonumber \\ &&
+ C_{b\uparrow} \cos(J_{\rm em}t/\hbar) e^{-\Gamma t/2} |X_b, \uparrow\rangle
+ C_{0\uparrow} \sqrt{1 - e^{-\Gamma t}} |0, \uparrow\rangle \nonumber \\ &&
+ C_{d\downarrow}\sin(J_{\rm em}t/\hbar) |X_d, \downarrow\rangle
+ C_{d\uparrow} |X_d, \uparrow\rangle,
\label{eq1_1}
\end{eqnarray}
with $C_{b\downarrow}=1/\sqrt{2}$.
Note that $|X_b, \uparrow\rangle$ and $|X_d, \downarrow\rangle$ coherently oscillate because $J_{\rm em}$ in off-diagonal elements of $H_{\rm xm}$ mix these two states.
Also from $\rho_{66}(\Gamma t >> 1) \rightarrow 1/2$ we deduce $|C_{0\uparrow}(\Gamma t >> 1)| \rightarrow 1/\sqrt{2}$, and finally $C_{d\uparrow}(t)=0$ because $\rho_{44}=0$.
The rest of coefficients in $|\psi(t)\rangle$ can be determined numerically by fitting to the solutions of density matrix that also fulfills the normalization of the wavefunction $\langle \psi |\psi\rangle = 1$.

The results obtained in this section illustrate that a bright-exciton transfers the angular-momentum of the CW laser field to the magnetization of the MI during the transient time, $t < t_r \approx 10 \Gamma^{-1}$. However, it gradually losses the magnetization to the environment via annihilation of the exciton within the  exciton annihilation time $t_r$.
Two vacuum states $|0,\uparrow\rangle$ and $|0,\downarrow\rangle$ are equally populated within $t_r$, hence the final MI magnetization is randomized and its ensemble average vanishes.
Note that the lack of DMP and build up of MI magnetization is the consequence of the ground state with random and uncorrelated states of MI.
If the annihilation of the exciton is selectively blocked for one type of spin of MI, e.g., by interruption of the decay process by quantum jumps within $t < t_r$, a dramatic change in the dynamics of the system occurs due to interaction between MI and other excitons in the environment and a final state with non-vanishing MI magnetization appears.
As seen in Fig.~\ref{fig1}, the time evolution of the density matrix predicts that the state of MI after first quantum jump ($t < t_r$) is partially spin-polarized. Thus the MI with  partial spin polarization interacts with second exciton tunneling in from the small quantum dot and as a result a net spin polarization builds up .
The rest of this paper is devoted to discussion of the DMP by quantum jumps.

\subsection{Quantum jump algorithm}
\label{sec3}
As noted in Ref.~\onlinecite{Hohenester2001:SSC}, detection of photons from a single quantum system requires spontaneous emission due to vacuum fluctuations, i.e., the photon emission is a stochastic process, described by quantum jump approach~\cite{Plenio1999:RMP,Hohenester2001:SSC}.
The time evolution of the density matrix of a   QD interacting with photons is given by Eq.(\ref{eq02}).
At zero temperature $n_B=0$ and the time evolution of the density matrix is described by a standard Lindblad master equation (ME)~\cite{Plenio1999:RMP}
\begin{equation}
\frac{d\rho}{dt}=\frac{-i}{\hbar}[H_{QD},\rho]-\frac{\Gamma}{2}\sum_{M_z} (\{P^\dagger_{M_z} P_{M_z},\rho\} -2 P_{M_z} \rho P^\dagger_{M_z}),
\label{eq3}
\end{equation}
where $\{\dots\}$ is the anti-commutator.
Comparing Eq.(\ref{eq3}) with Eq.(\ref{eq02}) we identify quantum jump operator $\hat{P}_{M_z} = |0,M_z\rangle \langle X_b,M_z|$ and its Hermitian conjugate $\hat{P}^\dagger_{M_z} = |X_b,M_z\rangle \langle 0,M_z|$ that project the excitonic state onto vacuum and vice versa without flipping spin of MI, hence
$\sum_{M_z}\hat{P}_{M_z}^\dagger \hat{P}_{M_z} = |X_b,\uparrow\rangle \langle X_b,\uparrow| + |X_b,\downarrow\rangle \langle X_b,\downarrow|$.
At each instance of time, $t$, the density matrix can be divided into series of density matrices, each representing a specific quantum trajectory associated with a sequence of randomly generated quantum jumps in the interval of time $[0,t]$. Hence the QD density matrix can be calculated by ensemble average of density matrices over all quantum jump trajectories.

The formulation of quantum jump starts from Eq.(\ref{eq3}). In the absence of MI, a recipe for quantum jump algorithm can be found in Ref.~\onlinecite{Plenio1999:RMP}. For completeness of our presentation  we first review this algorithm and then generalize it to exciton in the presence of MI and nuclear spins.
We consider optical transition in a two-level system consisting of bright-exciton and vacuum without considering an intermediate transition to dark-exciton. This condition is fulfilled if we disregard presence of any MI and nuclear spin.
Here the quantum jump operators are $\hat{P} = |0\rangle \langle X_b|$, $\hat{P}^\dagger = |X_b\rangle \langle 0|$,
hence $\hat{P}^\dagger \hat{P} = |X_b\rangle \langle X_b|$.
Starting at $t=0$ with the initial condition $|\psi(t=0)\rangle = |X_b\rangle$, we calculate the time evolution of the system in discrete time-steps $\delta t$. In each time-step we evaluate the quantum jump probability by calculating $\delta q_0 = \Gamma (\delta t) \langle \psi | \hat{P}^\dagger \hat{P} |\psi\rangle = \Gamma \delta t$ and drawing a random number $r$. If $r < \delta q_0$ a quantum jump occurs and $|\psi\rangle$ collapses to $|0\rangle$, otherwise $|\psi(t+\delta t)\rangle=e^{-\Gamma \delta t (\hat{P}^\dagger \hat{P})/2} |\psi(t)\rangle$. Here the generator for the time-evolution operator is a non-Hermitian Hamiltonian $H_{\rm eff} = - i \hbar \Gamma (\hat{P}^\dagger \hat{P})/2$.
So at $t=0 + \delta t$ we have $|\psi(0+\delta t)\rangle=e^{-\Gamma \delta t (|X_b\rangle \langle X_b|)/2} |X_b\rangle = e^{-\Gamma \delta t/2} |X_b\rangle + \sqrt{1 - e^{-\Gamma \delta t}} |0\rangle$. The last term keeps the norm of $|\psi\rangle$ constant (if we use the norm of wave-function as a constraint in our calculation). At this time $\delta q_1 = \Gamma (\delta t) e^{-\Gamma \delta t}$. We draw $r$ and if $r < \delta q_0 + \delta q_1$ then $|X_b\rangle \rightarrow |0\rangle$ and a photon is detected and calculation is terminated. Otherwise,
$|\psi(\delta t + \delta t)\rangle=e^{-\Gamma \delta t (|X_b\rangle \langle X_b|)/2} |\psi(0 + \delta t)\rangle = e^{-\Gamma (2\delta t)/2} |X_b\rangle + \sqrt{1 - e^{-\Gamma (2\delta t)}} |0\rangle$.
In $n$th-step $\delta q_n = \Gamma (\delta t) e^{- n \Gamma\delta t}$ thus
we calculate a cumulative quantum jump probability: 
\begin{eqnarray}
\delta p_n = \sum_{k=0}^n \Gamma (\delta t) e^{- k \Gamma\delta t}= \Gamma (\delta t) \frac{1 - e^{- (n+1) \Gamma\delta t}}{1 - e^{- \Gamma\delta t}},
\label{eq001}
\end{eqnarray}
and if $r < \delta p_n$ quantum jump occurs.
As the time advances, the chance for a quantum jump becomes more likely, however, the probability amplitude for $X_b$ in $|\psi\rangle$ decreases with the same rate simultaneously. In $n$th-step if there is still no quantum jump, then $|\psi(n \delta t)\rangle=e^{-\Gamma \delta t (|X_b\rangle \langle X_b|)/2} |\psi([n-1]\delta t)\rangle = e^{-\Gamma (n\delta t)/2} |X_b\rangle + \sqrt{1 - e^{-\Gamma (n\delta t)}} |0\rangle$.

The quantum jump algorithm in the presence of MI is similar to the one in the absence of MI, with a difference that the time evolution of the wavefunction is generated by an effective Hamiltonian $H_{\rm eff} = H_{QD} - i  \hbar \Gamma \delta t (\hat{P}^\dagger \hat{P})/2$ that allows an intermediate transition to the dark-exciton due to spin-exchange with MI.
Therefore the description of quantum jump process in the presence of MI is based on a three level system depicted in the inset of Fig.~\ref{fig3} and consist of $|0\rangle$, $|X_b\rangle$, $|X_d\rangle$ and MI.

\section{dynamical evolution of MI by train of excitons in the presence of nuclear spins}
\label{sec4}

As illustrated in Fig.1 a small quantum dot is continuously refilled by  a non-resonant circularly polarized CW laser.  The spin polarized excitons transfer into the QD containing the complex spin system MI.  We assume therefore a train of incoming bright excitons $|X_b\rangle\equiv|\downarrow,\Uparrow\rangle$ interacting with MI in the quantum dot. Each electron in the exciton transfers spin to MI and creates a superposition of dark and bright excitons entangled with MI and nuclear spins.  At the bright exciton recombination time, $t_r$, photon is detected, quantum jump takes place,  dark exciton wavefunction is erased and  MI and nuclear spin complex is left in a modified state.  The exciton removal is performed by using the quantum jump projector method~\cite{Plenio1999:RMP,Hohenester2001:SSC} described below which yields the modified wavefunction of the MI and nuclear spins. New spin polarized exciton tunnels into the quantum dot and begins interaction with the  MI and nuclear spins modified by electron spin of previous exciton.

The basis for combined exciton-spin system  is composed of three groups of basis states: vacuum $|0, M_z, I_{z1}, \dots, I_{zN_b}\rangle$, bright exciton $|X_b, M_z, I_{z1}, \dots, I_{zN_b}\rangle$ and dark exciton $|X_d, M_z, I_{z1}, \dots, I_{zN_b}\rangle$. Only the vacuum and bright exciton group of states are coupled to the photon field via  projectors $P_\lambda=|0,\lambda\rangle\langle X_b,\lambda|$. Here states $|\lambda\rangle=|M_z, I_{z1}, \dots, I_{zN_b}\rangle$ describe a total of $N_S=(2M_z+1)(2I_z+1)^{N_b}$ complex spin MI and nuclear spin states. In the following symbols $|\lambda\rangle$ and $|\mu\rangle$ represent $|M_z, I_{z1}, \dots, I_{zN_b}\rangle$.

The time evolution of the density matrix $\rho=|\Psi\rangle\langle\Psi|$ in ME, Eq.(\ref{eq3}) can be generalized by $M_z \rightarrow \lambda$. As described in section~\ref{sec3}, the wave-function $|\Psi\rangle$ subjected to stochastic ``birth-death" process~\cite{vanKampenbook} of recombination and photo-excitation can be used to describe the time propagation of the system coupled with radiation-field and undergoing quantum jump process.
At $t=0$ we start with the initial state $|\Psi^{n=0}(t=0) \rangle =|0\rangle|\chi_0\rangle$ of MI and the nuclear spin-bath. The index $n$ counts the number of quantum jump events.
The state $|\chi_0\rangle=\sum_{\lambda}C^{n=0}_{\lambda}|\lambda\rangle$ is a random  linear combination of all possible configurations with the coefficients $C^{(0)}_{\lambda}$ being uniformly distributed random complex numbers. We note that if we were to compute expectation value $\langle M_z\rangle$ for this random state we would obtain a finite value. However, averaging over many sets of
$C^{(0)}_{\lambda}$ yields no initial magnetization.

At $t=0^+$ a bright exciton created in neighboring QD enters the central QD. The creation of $|X_b\rangle$ and annihilation of $|0\rangle$ are described by operator $|X_b\rangle \langle 0|$, hence the wave-function of the system with one exciton is given by
\begin{eqnarray}
|&\Psi^{n=1}&(t=0^+)\rangle = |X_b\rangle \langle 0|\Psi^{n=0}(t=0)\rangle =  |X_b\rangle|\chi_0\rangle\nonumber \\ &&
= |X_b\rangle \sum_{\lambda} C^{n=0}_{\lambda}(t=0) |\lambda\rangle \nonumber \\ &&
= \sum_{\lambda} C^{n=1}_{\lambda}(t=0^+) |X_b,\lambda\rangle
\label{eq4}
\end{eqnarray}
where $C^{n=1}_{\lambda}(t=0^+)=C^{n=0}_{\lambda}(t=0)$.
The initial wave-function of the injected bright exciton, MI and nuclear-spins is an uncorrelated state.
However, the Hamiltonian $H_{QD}$ that accounts for the exchange coupling, creates quantum correlation in the exciton-MI-nuclei complex and $|\Psi\rangle$ evolves into a linear combination of all configurations, including an entangled state between bright and dark excitons.
As a function of time, the bright-exciton decays into vacuum because of coupling with quantized electromagnetic-field.

To be consistent with the quantum jump algorithm we discretize the time $t$ into small steps $\delta t$. Note that because of the small eh- and MI-nuclear-spin couplings ($J_{\rm ne}$, $J_{\rm nh}$ and $J_{\rm nh}$) the excitons and MI evolve in a frozen-fluctuating field of nuclear-spins~\cite{Merkulov2002:PRB,Melikidze2004:PRB,Abolfath2010:PRB}. The eh recombination time, $t_r$ is the smallest time-scale in our model, hence $\delta t << t_r$.

The time evolution of the wave-function of eh-MI-nuclei complex is calculated in the Schr\"odinger picture~\cite{Melikidze2004:PRB} by using the relation $|\Psi(t+\delta t)\rangle=\exp(-iH_{\rm eff}\delta t/\hbar)|\Psi(t)\rangle$. Here, $|\Psi(t)\rangle$ is the wave-function of the entire system, $H_{\rm eff} = H_{QD} - i\hbar(\Gamma/2)\sum_{\lambda}P^\dagger_{\lambda}P_{\lambda}$ where the last term describes the decay of bright exciton due to coupling with photon-field, hence
\begin{eqnarray}
|\Psi^{n=1}(\delta t)\rangle = \exp(-i H_{\rm eff}\delta t/\hbar) |\Psi^{n=1}(t=0^+)\rangle.
\label{eq5}
\end{eqnarray}
In Eq.~\ref{eq5} we use $\exp(-i H_{\rm eff} \delta t/\hbar) \approx \exp(-i H_{QD}\delta t/\hbar) \exp (- \Gamma\delta t/2 \sum_{\lambda}P^\dagger_{\lambda}P_{\lambda}) + O((\delta t)^2)$ with
$P_\lambda^\dagger P_\lambda = |X_b, \lambda\rangle \langle X_b, \lambda|$ and the following identities
$\exp (- \Gamma\delta t \sum_\lambda P_\lambda^\dagger P_\lambda)  |X_b,\mu\rangle = \prod_\lambda\exp (- \Gamma \delta t P_\lambda^\dagger P_\lambda)  |X_b, \mu\rangle = \exp (- \Gamma \delta t)  |X_b, \mu\rangle$,
as $\exp (- \Gamma\delta t P_\lambda^\dagger P_\lambda)  |X_b,\lambda\rangle = \exp (- \Gamma \delta t) |X_b, \lambda\rangle$, and
$\exp (- \Gamma\delta t P_\lambda^\dagger P_\lambda)  |X_b,\mu\rangle = |X_b, \mu\rangle$ where $\mu\neq \lambda$, as well as
$\exp (- \Gamma\delta t P_\lambda^\dagger P_\lambda) |X_d, \lambda\rangle = |X_d, \lambda\rangle$, and
$\exp (- \Gamma\delta t P_\lambda^\dagger P_\lambda) |0, \lambda\rangle = |0, \lambda\rangle$.

Because $H_{\rm eff}$ is time-independent, we employ the method based on Bessel-Chebyshev polynomial expansion~\cite{Melikidze2004:PRB} to calculate the time evolution of the wave-function
\begin{eqnarray}
&&|\Psi^{n=1}(\delta t)\rangle \approx  e^{- \frac{\Gamma}{2}\delta t \sum_{\lambda}P^\dagger_{\lambda}P_{\lambda}}
e^{-\frac{i}{\hbar}H_{QD}\delta t} |\Psi^{n=1}(t=0^+)\rangle \nonumber \\ &&
= e^{- \frac{\Gamma}{2}\delta t \sum_{\lambda}P^\dagger_{\lambda}P_{\lambda}} \sum_{\mu} [\tilde{C}^{n=1}_{X_b,\mu}(\delta t) |X_b,\mu\rangle+ \nonumber\\ &&
+ \tilde{C}^{n=1}_{X_d,\mu}(\delta t) |X_d,\mu\rangle].
\label{eq08}
\end{eqnarray}
Note that $H_{QD}$ is Hermitian and thus $\exp(-i H_{QD}\delta t/\hbar)$ is a unitary operator that conserves the norm of wave-function, hence $|\tilde{C}^{n=1}_{X_b,\mu}(\delta t)|^2+|\tilde{C}^{n=1}_{X_d,\mu}(\delta t)|^2=1$. This is in contrast with the operator $\exp (- \Gamma \delta t/2 \sum_\lambda P_\lambda^\dagger P_\lambda)$ that is non-unitary and does not preserve norm of wave-function, however, because it describes the decay of $X_b$ into vacuum, we build a norm-conserving wave-function by adding vacuum.
After applying $\exp (- \Gamma \delta t/2 \sum_\lambda P_\lambda^\dagger P_\lambda)$ in Eq.~\ref{eq08} we find
\begin{eqnarray}
&&|\Psi^{n=1}(\delta t)\rangle = \sum_{\lambda} |\lambda\rangle
[\exp(-\Gamma \delta t/2) C^{n=1}_{X_b,\lambda}(\delta t) |X_b\rangle \nonumber \\ &&
+ \sqrt{1-\exp(-\Gamma \delta t)} C^{n=1}_{0,\lambda}(\delta t) |0\rangle \nonumber \\ &&
+ C^{n=1}_{X_d,\lambda}(\delta t) |X_d\rangle].
\label{eq8}
\end{eqnarray}
In the limit of $\Gamma=0$ the coefficients with and without tilde used in Eqs.~\ref{eq08}-\ref{eq8} are identical.
Matching the initial conditions between Eq.(\ref{eq4}) and Eq.(\ref{eq8}) implies $C^{n=1}_{X_b,\lambda}(\delta t=0) = C^{n=1}_{\lambda}(t=0^+)$ and $C^{n=1}_{X_d,\lambda}(\delta t=0)=0$.
Using an iterative procedure to propagate the wavefunction in time we find
\begin{eqnarray}
&&|\Psi^{n=1}(t)\rangle =
\sum_{\lambda} |\lambda\rangle
[C^{n=1}_{X_b, \lambda}(t) |X_b\rangle
+ C^{n=1}_{0, \lambda}(t) |0 \rangle \nonumber \\ &&
+ C^{n=1}_{X_d, \lambda}(t) |X_d\rangle],
\label{eq9}
\end{eqnarray}
where the coefficients $C^{n=1}_{X_b, \lambda}(t)$, $C^{n=1}_{0, \lambda}(t)$, and $C^{n=1}_{X_d, \lambda}(t)$ are determined numerically.
This wave-function describes a correlated state of bright and dark excitons as well as vacuum.
Because of the spin flip-flop process of electron with MI and nuclear-spins the initially formed bright exciton $|X_b\rangle$ mixes with the dark-exciton $|X_d \rangle$.

From the Lindblad ME, the quantum jump transition rate is given by $\Gamma_{\rm jump}=\Gamma\rho_{b}$. $\rho_{b}$ represents the population of the bright exciton obtained from the full QD density matrix after tracing over MI and nuclear spin degrees of freedom.
The quantum jump probability $\delta p_{\rm jump} = \Gamma_{\rm jump} \delta t$ is then calculated and compared with a random number $r$ generated between zero and one. If $r < \int_0^{t_r}dt  \Gamma_{\rm jump}$ a quantum jump takes place, photon is recorded and the quantum dot is in the ground state. The elapsed time $t_r$ recorded for this quantum jump is the eh-recombination time.
At $t=t_r$ we allow exciton to annihilate by spontaneous emission of a photon. The operator that allows annihilation of bright exciton and creation of vacuum is $|0\rangle \langle X_b|$. Hence
\begin{eqnarray}
|\Psi^{n=2}(t=t_r)\rangle &=& |0\rangle \langle X_b| \Psi^{n=1}(t=t_r)\rangle \nonumber \\ &&
= |0\rangle \sum_{\lambda} C^{n=1}_{X_b,\lambda}(t=t_r) |\lambda\rangle \nonumber \\ &&
= \sum_{\lambda} C^{n=2}_{\lambda}(t=t_r) |0,\lambda\rangle
\end{eqnarray}
where $C^{n=2}_{\lambda}(t=t_r)=C^{n=1}_{X_b,\lambda}(t=t_r)$.

Immediately after annihilation of exciton, a new spin polarized exciton tunnels into the quantum dot from the neighboring dot. The spin polarized exciton interacts with the MI spin $M$ and nuclear spins $I$ in a state modified by the previous exciton.
At $t=t_r + 0^+$, second bright exciton $X_b$ created in the neighboring QD tunnels into the central QD
\begin{eqnarray}
&&|\Psi^{n=3}(t=t_r + 0^+)\rangle = |X_b\rangle \langle 0|\Psi^{n=2}(t=t_r)\rangle \nonumber \\ &&
= |X_b\rangle \sum_{\lambda} C^{n=2}_{\lambda}(t=t_r) |\lambda\rangle \nonumber \\ &&
= \sum_{\lambda} C^{n=3}_{\lambda}(t=t_r + 0^+) |X_b,\lambda\rangle
\end{eqnarray}
with matching the initial conditions that requires $C^{n=3}_{\lambda}(t=t_r + 0^+)=C^{n=2}_{\lambda}(t=t_r)$.
Note that this state is not correlated. The quantum correlation appears from the time evolution of wave-function generated by exchange couplings in $H_{\rm eff}$ right after $t=t_r$
\begin{eqnarray}
|\Psi^{n=3}(t)\rangle &=& \exp(-iH_{\rm eff}t/\hbar) |\Psi^{n=3}(t=t_r+0^+)\rangle \nonumber \\ &&
= \sum_{\lambda} |\lambda\rangle
[C^{n=3}_{X_b,\lambda}(t) |X_b\rangle + C^{n=3}_{0,\lambda}(t) |0\rangle \nonumber \\ &&
+ C^{n=3}_{X_d,\lambda}(t) |X_d\rangle],
\end{eqnarray}
where $C^{n=3}_{X_b,\lambda}(t_r + 0^+)=C^{n=3}_{\lambda}(t_r  + 0^+)$ and $C^{n=3}_{X_d,\lambda}(t_r + 0^+)=0$ are matching conditions.
As we see there is no type of linear combination between $|0\rangle$ and $\{|X_b\rangle, |X_d\rangle\}$ because there is no Rabi-oscillations between vacuum and excitons.

To summarize the above procedure and make connection between tunneling of exciton and photo-emission we formally introduce a projector $Q_{n\rightarrow n+1}=|X^{n+1}_b\rangle\langle X^{n}_b|$ in $n$th step of quantum jump. The superscripts refer to annihilated $n$th and created $n+1$th exciton.
Note that $Q_{n\rightarrow n+1}|X^{n}_b\rangle=|X^{n+1}_b\rangle$ and $Q_{n\rightarrow n+1}|X_d\rangle=Q_{n\rightarrow n+1}|0\rangle=0$,
hence the quantum jump operator projects out any correlated state composed of superposition of bright and dark exciton to a new born bright exciton.
The new  wavefunction in the QD then can be constructed as $|\Psi(t=t_r^+)\rangle=Q_{n\rightarrow n+1}|\Psi(t=t_r^-)\rangle$ where $t_r^\pm = t_r \pm \eta$ and $\eta \rightarrow 0$. In this state, $|X^{n+1}_b\rangle$ is initially uncorrelated from MI and nuclear-spins. At $t=t_r^+$ it can be expressed as \begin{equation}
|\Psi^{n+1}\rangle = |X^{n+1}_b \rangle \sum_{\lambda} A_N C^{n}_{X_b, \lambda} (t_r )|\lambda\rangle.
\end{equation}
We observe that after quantum jump the new injected exciton $X_b^{n+1}$ starts with the normalized (factor $A_N$) state of MI and nuclear spins
$\sum_{\lambda} C^{n}_{X_b, \lambda} (t_r )|\lambda\rangle$ which was left over by the previous bright exciton $X_b^{n}$ at the time of radiative recombination. Detecting a photon erased the dark exciton wave-function and modified the state of both MI and nuclear spins. This is the DMP mechanism discussed here.
With initial condition established, the time evolution of the  entangled state of photo-carriers with MI and spin-bath then can be calculated after updating
the coefficients $C$'s. At the end one needs to average over initial conditions.
Although the procedure discussed here describes the immediate refilling of central QD after annihilation of the exciton, we can always implement a waiting time between the recombination and refilling process.

\section{Numerical results and discussion}
\label{sec5}
Our approach to DMP is illustrated using parameters based on (Cd,Mn)Te QDs with $\tilde{J}_{\rm em}=15$ meV nm$^3$, $\tilde{J}_{\rm hm}=60$ meV nm$^3$ corresponding to the exchange coupling in the bulk materials, hence $J_{\rm em}=\tilde{J}_{\rm em} |\phi_{\rm e}(R_m)|^2$ and $J_{\rm hm}=\tilde{J}_{\rm hm} |\phi_{\rm h}(R_m)|^2$. The circular symmetry of quantum dots is implemented by assuming $\Delta_1=0$.
Here $\phi_{\rm e/h}(\vec{R}_{\rm m})$ is the e/h envelope-wavefunction in the central dot at $\vec{R}_{\rm m}$, the position of MI.
We assume $J_{\rm eh}=0.6$ meV~\cite{Bayer2002:PRB} and initialize $J_{\rm ne}$, $J_{\rm nh}$, $J_{\rm nm}$ and $J_{\rm nn'}$ as random numbers with a mean value of the order of 1 $\mu$eV. However we note that the realistic value for nuclear hyperfine interaction is reported within 1 neV~\cite{Coish2009:PSSB} three orders of magnitude smaller than the energy scales used in our finite size calculation.

\begin{figure}
\begin{center}
\vspace{1cm}
\includegraphics[width=0.98\linewidth]{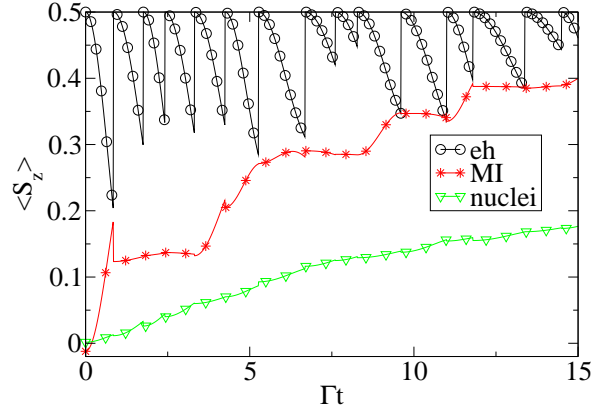}
\caption{
Time evolution of $S_z$ of a train of injected photo-electrons inside QD (circles), a single MI (stars) and the average of $N_b=15$ nuclear-spin polarization (triangles).
}
\label{fig4}
\end{center}
\end{figure}

\begin{figure}
\begin{center}
\vspace{1cm}
\includegraphics[width=0.98\linewidth]{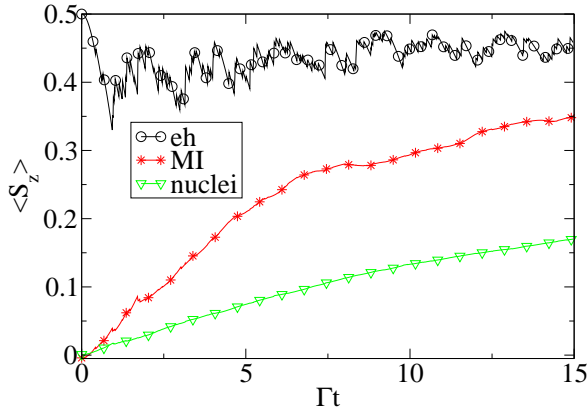}
\caption{
Time evolution of ensemble average of 20 trajectories for $S_z$ of a train of injected photo-electrons inside QD (circles), a single MI (stars) and the average of $N_b=15$ nuclear-spin polarization (triangles).
}
\label{fig5}
\end{center}
\end{figure}

\begin{figure}
\begin{center}
\vspace{1cm}
\includegraphics[width=0.98\linewidth]{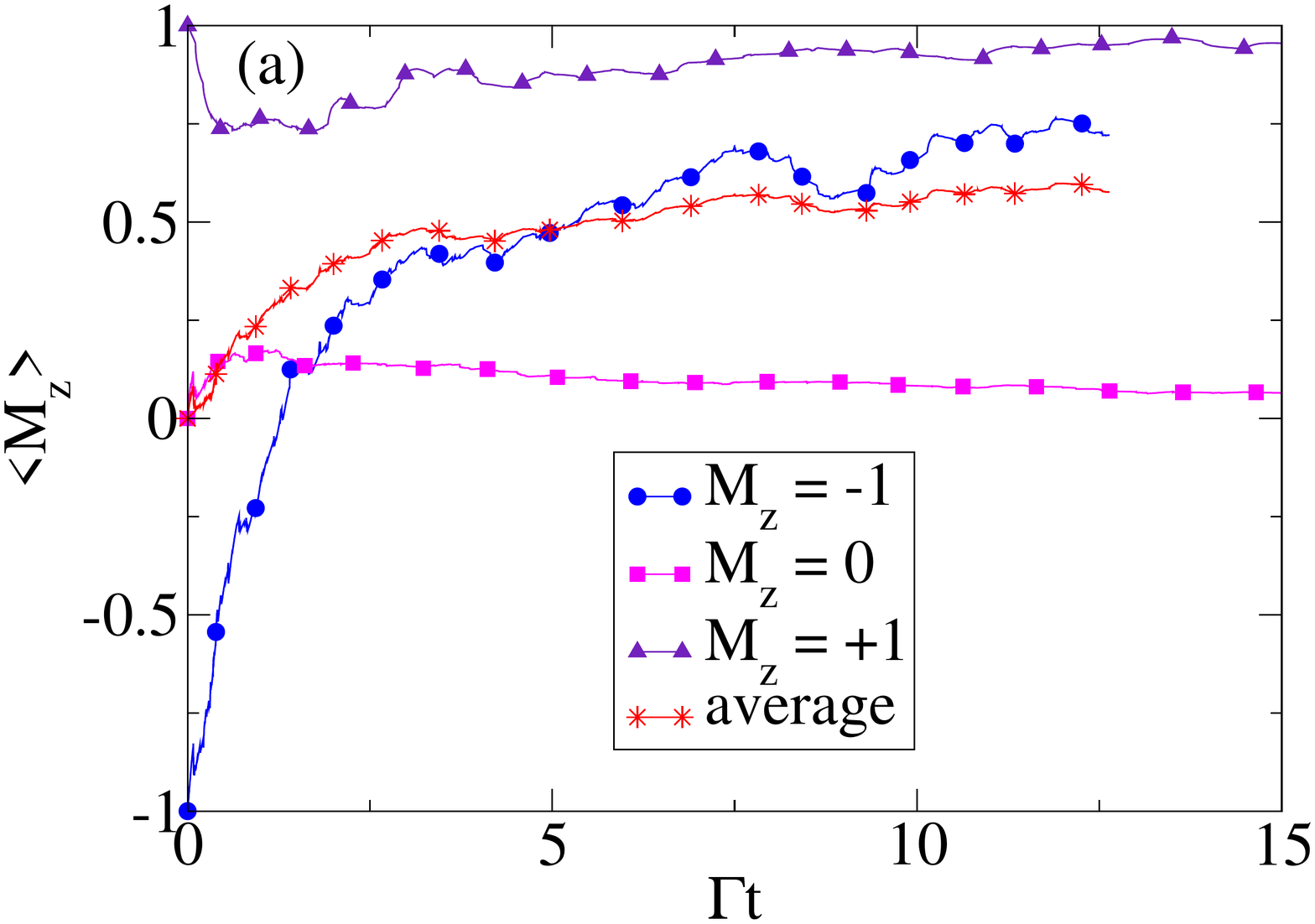} \\
\includegraphics[width=0.98\linewidth]{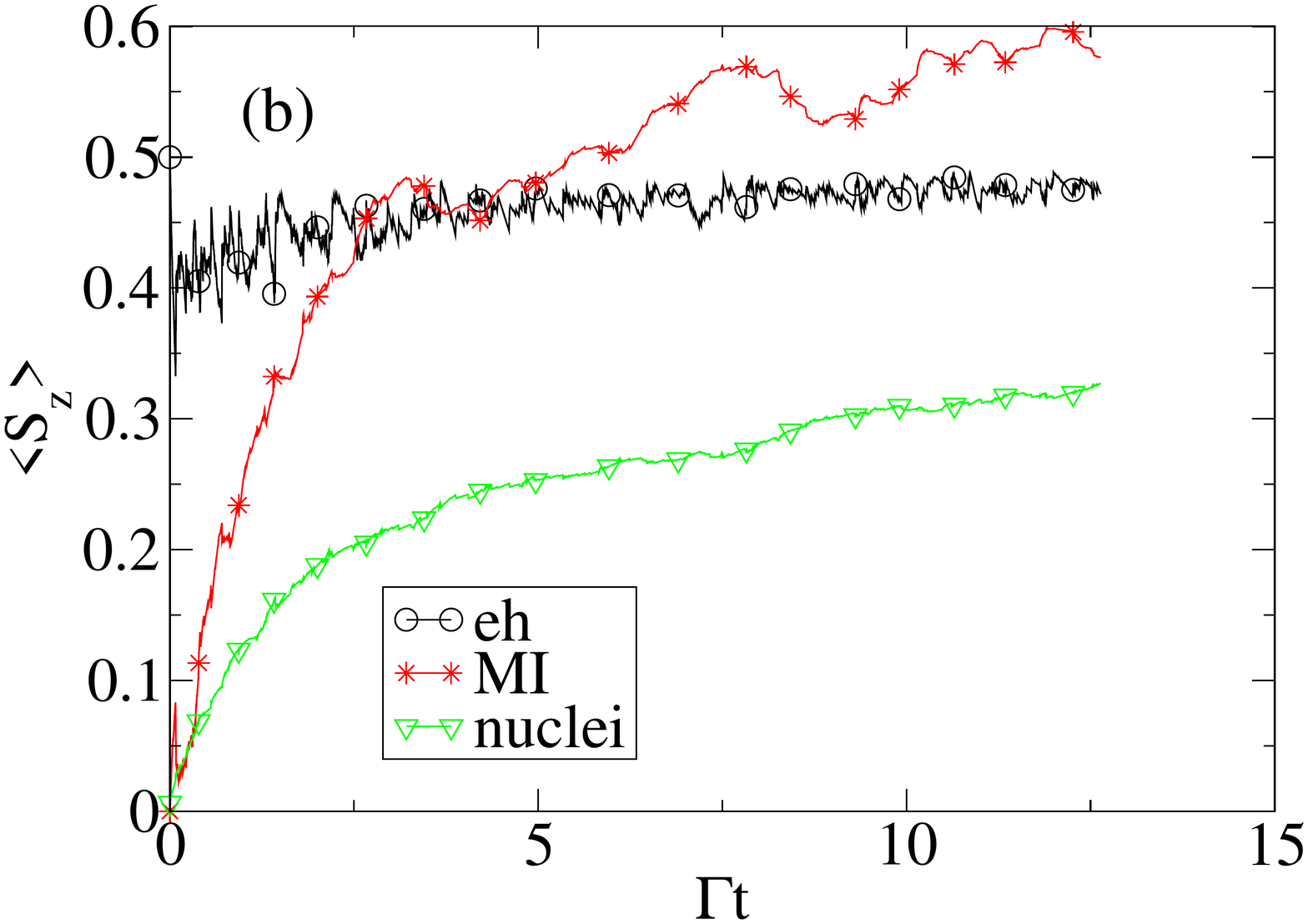}
\caption{
(a) Time evolution of magnetic moment of ensemble of MIs with two localized p/d-electrons in ferromagnetically ordered spin-triplet $M=1$ and $M_z=\pm 1, 0$ interacting with a train of injected photo-electrons inside QD, and $N_b=15$ nuclear-spins. Unlike $M_z=\pm 1$ that interact strongly with excitons, $M_z = 0$ exhibits weak interaction. The average of three states show saturation close to $\langle M_z \rangle = \frac{1}{3}(2\times 1 + 0) = 0.67$ as the ensemble is populated equally among all possible spin-triplet states.
(b) Time evolution of ensemble averaged of spin of MI (stars), photo-electrons (circles), and $N_b=15$ nuclear-spins (triangles).
}
\label{fig6}
\end{center}
\end{figure}

\begin{figure}
\begin{center}
\vspace{1cm}
\includegraphics[width=0.98\linewidth]{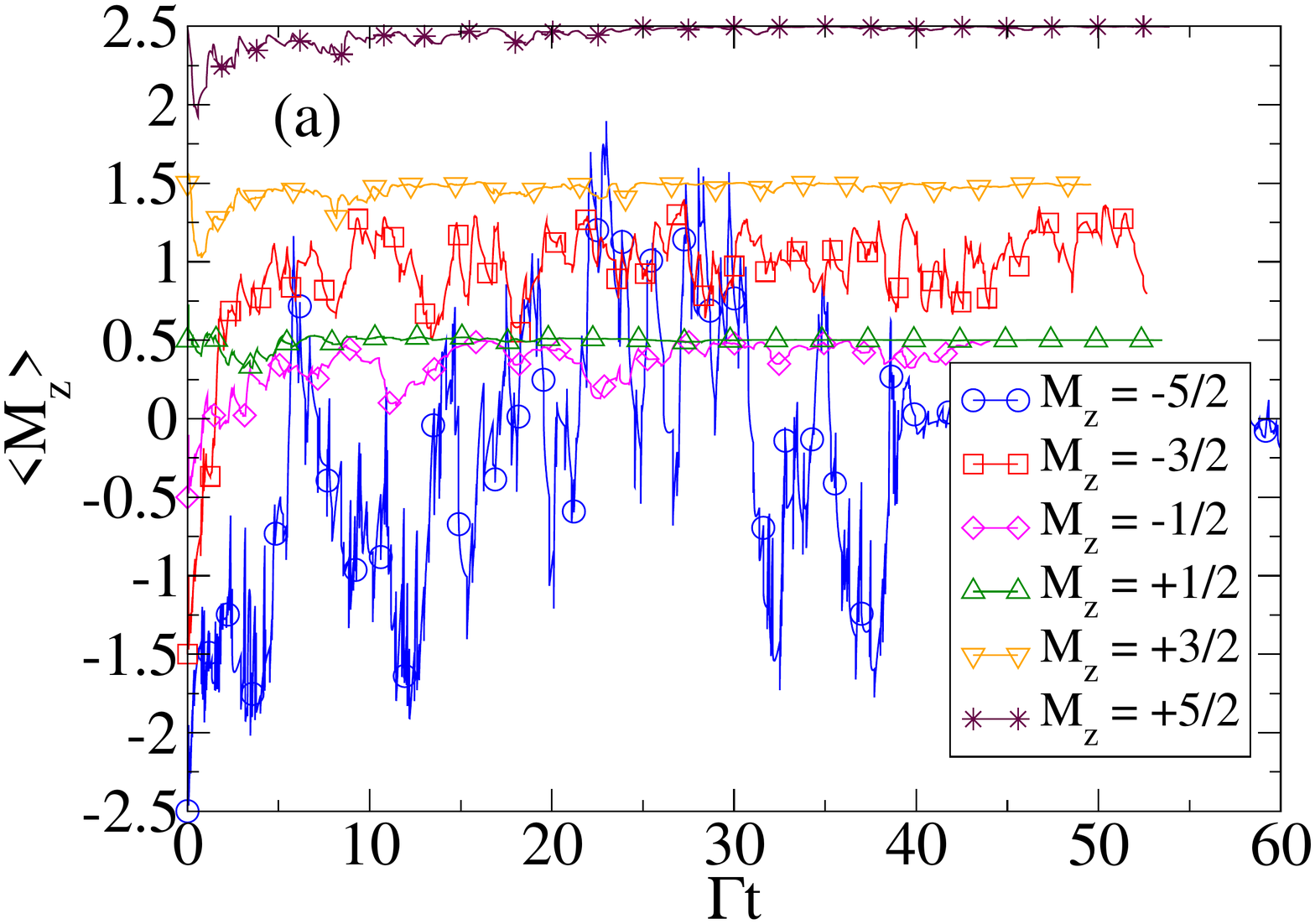} \\
\includegraphics[width=0.98\linewidth]{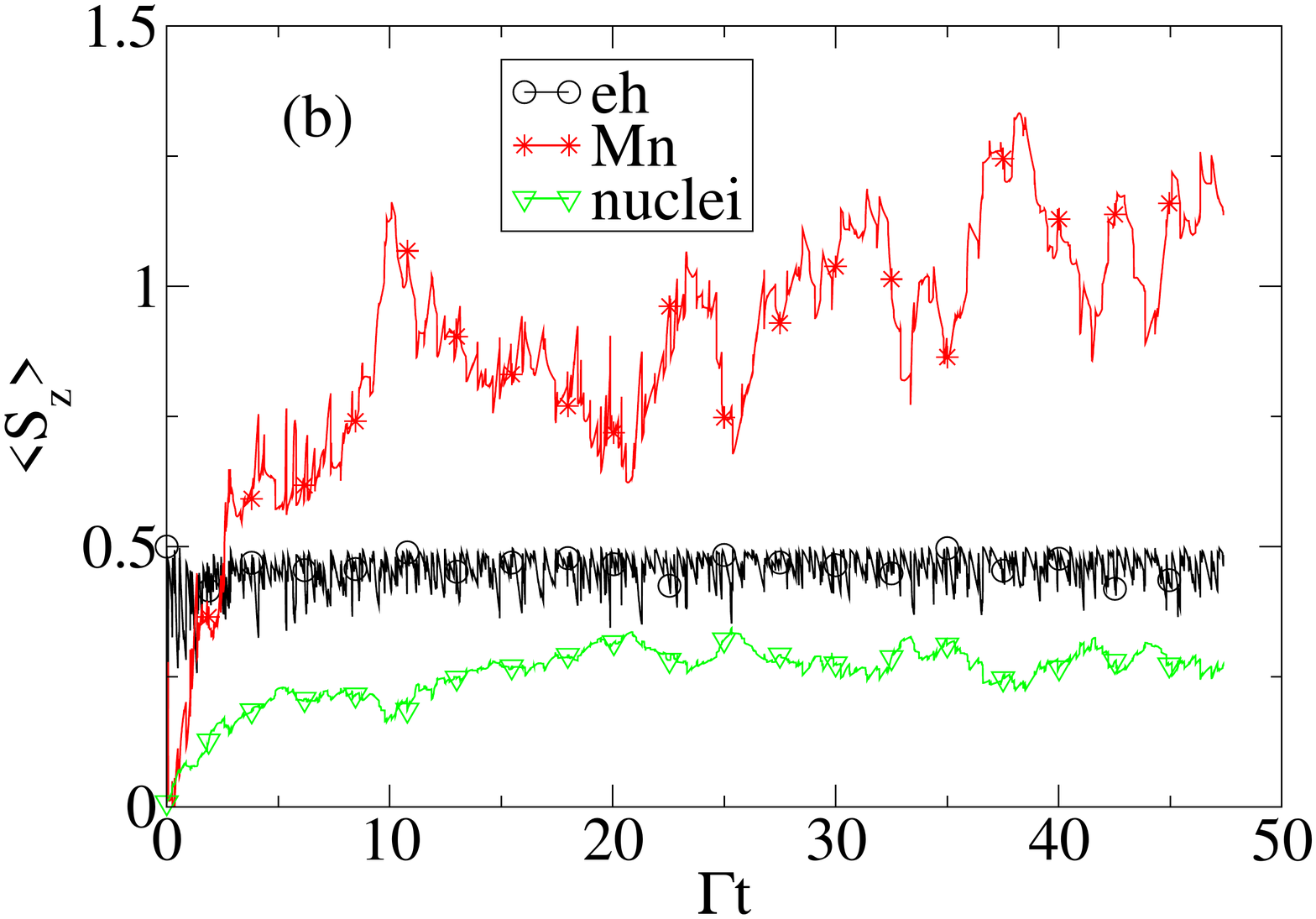}
\caption{
Time evolution of magnetic moment of individual states of Mn (a) and an ensemble of Mn's (b) with five localized d-electrons (stars) interacting with a train of injected photo-electrons inside QD (circles), and $N_b=15$ nuclear-spins (triangles).
The ensemble is constructed with equally populated $M_z=\pm5/2,\pm3/2,\pm1/2$ among finite number of Mn's.
The average of six states show saturation close to $\langle M_z \rangle = \frac{1}{6}(2\times\frac{5}{2} + 2\times\frac{3}{2}+ 2\times\frac{1}{2}) = \frac{9}{6} = 1.5$ as the ensemble is populated equally among all possible spin-triplet states.
}
\label{fig7}
\end{center}
\end{figure}


Here we discuss numerical results with nuclear spins, immediately after refilling of QD by bright exciton.
Figs.~\ref{fig4}-\ref{fig7} illustrate DMP/DNP and the quantum jump trajectories for exciton, MI and nuclear-spins.
In Fig.~\ref{fig4} a single quantum jump trajectory for spin-1/2 MI is plotted.
In Figs.~\ref{fig5}-\ref{fig7} the ensemble average of twenty quantum jump trajectories for spin-1/2 (Fig.~\ref{fig5}), spin-1 (Fig.~\ref{fig6}) and spin-5/2 (Fig.~\ref{fig7}) MI are plotted.
The trajectories are time-evolution of the initial spin wavefunctions which are generated in random linear combination of spin configurations.
Each curve consists of thousands of time-steps and points. For clarity of the legends, after every hundred points symbols like circle, star and triangle are superimposed on each curve.
As shown the MI and the average polarization of $N_b=15$ nuclear spins gradually builds-up by a train of injected bright excitons.
At $t=t_r$, one pair of eh collapses into vacuum with $\langle S_{{\rm e},z}(t) \rangle < 1/2$ as part of the e-spin is transferred to MI.
An empty dot instantaneously absorbs the second photo-generated eh pair with total angular momentum $j_z = -1$ which transfers to spin of MI and nuclei before its removal. We repeat this procedure until the spin polarization of MI and nuclear-spins is built-up.
The method presented here is limited to a finite number of nuclear spins because of exponentially increasing computational effort with the number of spins.
However, a systematic study of the convergence of the numerical results by increasing $N_b$ shows satisfactory outcomes around $N_b=15$.

We now discuss DMP for MIs with more than one localized electron. We consider two cases of MI with two and five electrons localized in open-shell p/d-orbitals.
The spin Hund's rule implies that the total spin of the electronic ground state of MIs is maximum. For two electrons the spin triplet manifold ($M=1$) is separated from the higher energy spin-singlet state ($M=0$) with the singlet-triplet energy gap $E_{M=0} - E_{M=1} = |J_m|$. Here $J_m$ is the ferromagnetic exchange coupling between two electrons localized in MI. Similarly the lowest energy state of MI (e.g., Mn) with five d-electrons corresponds to total spin $M=5/2$ with six-fold degeneracy. These states are separated from higher energy spin-manifold with energy gap proportional to $J_m$. Considering $J_m$ few times larger than other exchange couplings avoids mixing ground state with the higher energy excited states of MI.

For a system containing spin-1 MI, we consider an ensemble with equally populated states $M_z=\pm 1, 0$ (1/3 for each $M_z$). Similarly the ensemble of Mn's contains $M_z = \pm 5/2, \pm 3/2, \pm 1/2$ with equal population (1/6 for each $M_z$).
Thus ensemble average over all possible QJ trajectories includes summation over all $S_z$.
In Fig.~\ref{fig6}(a), we show the time evolution of spin-triplet states initially started from $M_z=\pm 1, 0$.
We observe that MI with $M_z=-1$ switches polarization to $\langle M_z \rangle \approx +1$ because of strong interaction with excitons that allows DMP mechanism to proceed efficiently. It is therefore expected that the polarization of $M_z(t=0)=+1$ does not alter dramatically, although it is initially decohered by nuclear spins but we find that it stays polarized with $\langle M_z \rangle \approx +1$ because of strong interaction with train of excitons.
On the other hand, MI with initial polarization $M_z(t=0)=0$ fluctuates around $M_z=0$. Analogue to its spin-singlet counter part, the spin-triplet $M_z=0$ weakly interacts with excitons and nuclear spins~\cite{Abolfath2012:PRL}.
Hence the the polarization obtained for this ensemble indicates that the final state is a mixture of all spin-triplet configurations with maximum achievable polarization $\langle M_z \rangle = \frac{1}{3}(2\times 1+0) = \frac{2}{3} = 0.67$ as $M_z = \pm 1$ equally contribute to ensemble average of $\langle M_z \rangle$.
Similarly we can predict the maximum spin polarization achievable for ensemble of Mn can be calculated by
$\langle M_z \rangle = \frac{1}{6}(2\times\frac{5}{2} + 2\times\frac{3}{2}+ 2\times\frac{1}{2}) = \frac{9}{6} = 1.5$ as shown in Fig.~\ref{fig7}. It is straightforward to show that mixing with higher energy excited states suppress the magnetic saturation down to $\langle M_z \rangle = 1.1$ if the final state is a mixture of equally populated all spin multiplicities of five spin-1/2 electrons. The results shown in Fig.~\ref{fig7} suggest that the saturation of $\langle M_z \rangle$ occurred between 1.1 and 1.5 that might be interpreted as an indication of leakage of the optically pumped ground state of Mn to its excited states.

\section{Summary}
\label{sec6}
In conclusion, dynamical magnetic and/or nuclear polarization in single quantum complex spin systems is discussed for the case of spin transfer from exciton to the central spin of magnetic impurity in a quantum dot in the presence of a finite number of nuclear spins. The exciton is described in terms of the electron and heavy hole spins interacting with  magnetic impurity via exchange interaction, with a finite number of nuclear spins via hyperfine interaction and with photons via dipole interaction.  The  time-evolution of the exciton, magnetic impurity and nuclear spins is calculated exactly between quantum jumps corresponding to exciton radiative recombination.
The collapse of the wave-function and the refilling of the quantum dot with new spin polarized exciton is shown to lead to a build up of magnetization of the magnetic impurity as well as nuclear spins. The competition between electron spin transfer to magnetic impurity and to nuclear spins simultaneous with the creation of dark excitons is therefore elucidated. The technique presented here opens up the possibility of studying optically induced Dynamical Magnetic Polarization in Complex Spin Systems.

\section{acknowledgement}
The authors thank NSERC, NRC and Canadian Institute for Advanced Research for support and hospitality. RMA thanks Steven Girvin for useful discussion and the support from Texas Advanced Computing Center (TACC) for computer resources.


$\dagger$ Present address: Department of Therapeutic Radiology, Yale School of Medicine, Yale University, New Haven, CT 06520



\end{document}